\documentclass[12pt,a4paper]{article}
\pdfoutput=1

\usepackage[english]{babel} 
\usepackage[utf8]{inputenc}
\usepackage[left=01in,right=1in,top=1.0in,bottom=1.1in]{geometry}
\usepackage[labelfont=bf]{caption}

\usepackage{dsfont}
\usepackage{amsfonts}
\usepackage{amsmath}
\usepackage{amssymb}
\usepackage{amsthm}
\usepackage{bbold}
\usepackage{calc}
\usepackage{cancel}
\usepackage{float}
\usepackage{slashed}
\usepackage{mathrsfs} 
\usepackage{pdfpages}
\usepackage{hyperref}
\usepackage{multirow}
\usepackage{mathtools}
\usepackage{braket}
\usepackage{cite}
\usepackage{tikz}
\usepackage{tikzsymbols}
\usepackage{array}
\usepackage{colortbl} 	
\definecolor{Gray}{gray}{0.95}
\definecolor{RGray}{gray}{0.90}
\definecolor{CGray}{gray}{0.92}
\usepackage{arydshln}
\usepackage{soul}
\usepackage{booktabs}
\usepackage{lipsum}
\usepackage[normalem]{ulem}

\numberwithin{equation}{section}
\numberwithin{figure}{section}
\numberwithin{table}{section}

\allowdisplaybreaks

\newcommand{\be}{\begin{equation}}
\newcommand{\ee}{\end{equation}}

\newcommand{\gsim}{\lower.7ex\hbox{$\;\stackrel{\textstyle>}{\sim}\;$}}
\newcommand{\lsim}{\lower.7ex\hbox{$\;\stackrel{\textstyle<}{\sim}\;$}}

\makeatletter
\g@addto@macro\bfseries{\boldmath}
\makeatother

\newcommand{\bea}{\begin{eqnarray}}
\newcommand{\eea}{\end{eqnarray}}

\begin{document}

\begin{center}
\vspace{0.7cm}
{\Large\bf Triple-leptoquark interactions for tree- and \\[0.22em] loop-level proton decays}\\[1.0cm] 
{Ilja Dor\v sner${}^{(a,b)}$,~Svjetlana Fajfer${}^{(b,c)}$}, 
{Olcyr~Sumensari${}^{(d)}$}\\
\vspace{0.7cm}

{\em\small ${}^{(a)}$University of Split, Faculty of Electrical Engineering, Mechanical Engineering and Naval Architecture (FESB), Ru\dj era Bo\v skovi\' ca 32, 21000 Split, Croatia\\[0.3em]
${}^{(b)}$J.\ Stefan Institute, Jamova 39, P.\ O.\ Box 3000, 1001 Ljubljana, Slovenia\\[0.3em]
${}^{(c)}$Department of Physics, University of Ljubljana, Jadranska 19, 1000 Ljubljana, Slovenia\\[0.3em]
${}^{(d)}$Universit\'e Paris-Saclay, CNRS/IN2P3, IJCLab, 91405 Orsay, France}
\end{center}
\vspace{0.5 cm}

\centerline{\large\bf Abstract}
\vspace{0.6em}
{\small We study the impact of triple-leptoquark interactions on matter stability for two specific proton decay topologies that arise at the tree- and one-loop level if and when they coexist. We demonstrate that the one-loop level topology is much more relevant than the tree-level one when it comes to the proton decay signatures despite the usual loop-suppression factor. We subsequently present detailed analysis of the triple-leptoquark interaction effects on the proton stability within one representative scenario to support our claim, where the scenario in question simultaneously features a tree-level topology that yields three-body proton decay $p\to e^+ e^+ e^-$ and a one-loop level topology that induces two-body proton decays $p\to \pi^0 e^+$ and $p\to \pi^+ \bar{\nu}$. We also provide a comprehensive list of the leading-order proton decay channels for all non-trivial cubic and quartic contractions involving three scalar leptoquark multiplets that generate triple-leptoquark interactions of our interest, where in the latter case one of the scalar multiplets is the Standard Model Higgs doublet.}
\thispagestyle{empty}
\setcounter{page}{0}


\section{Introduction}
\label{sec:introduction}   

If one adds to the Standard Model (SM) particle content more than one scalar leptoquark multiplet, the mixing between two or more electric charge eigenstates from these multiplets can generate interesting physical phenomena~\cite{Hirsch:1996qy}. For example, an explicit mixing between two specific leptoquark multiplets that is induced through a coupling with the Higgs boson can generate neutrino masses~\cite{Chua:1999si,Mahanta:1999xd,AristizabalSierra:2007nf,Babu:2010vp,Pas:2015hca,Dorsner:2017wwn} or contribute towards the magnetic and dipole moments of charged leptons~\cite{Dorsner:2019itg}. Moreover, the hypothetical presence of two scalar leptoquark states can provide an explanation of the observed discrepancies between the experimental measurements of charged and neutral-current semileptonic $B$-meson decays and the corresponding SM predictions~\cite{Crivellin:2017zlb,Angelescu:2018tyl,Angelescu:2021lln,Becirevic:2018afm,Crivellin:2019dwb,Gherardi:2020qhc,Saad:2020ihm}.

\begin{figure}[h!]
\centering
\includegraphics[width=0.85\textwidth]{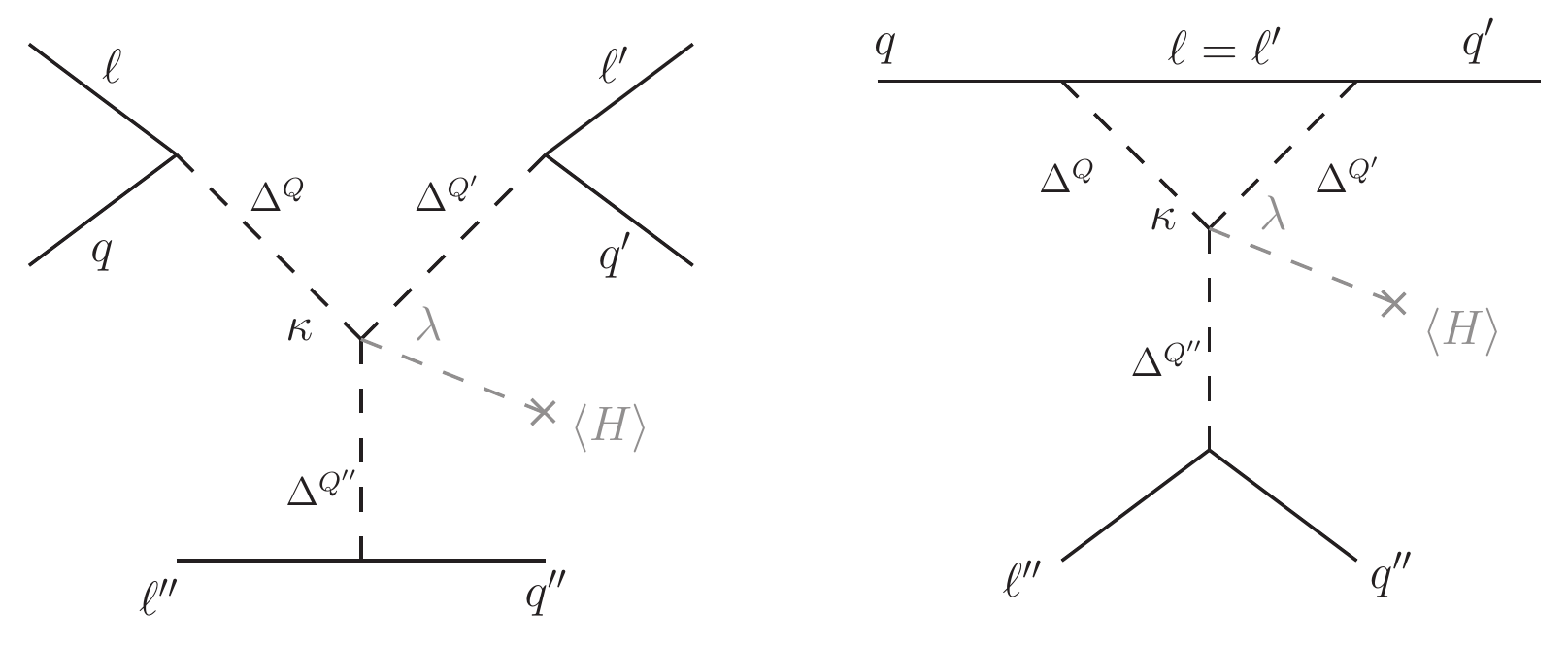}
\caption{Two different proton decay topologies generated by the triple-leptoquark interactions. Both can be with or without a Higgs vacuum expectation value leg insertion. $q$'s and $\ell$'s denote generic quarks and leptons of the SM while $\Delta^Q$, $\Delta^{Q^\prime}$, and $\Delta^{Q^{\prime\prime}}$ are scalar leptoquark mass eigenstates with electric charges $Q$, $Q^\prime$, and $Q^{\prime\prime}$, respectively. }
\label{fig:topologies}
\end{figure}

In this work we study one particular class of proton decay processes that requires existence of either two or three scalar leptoquark multiplets in addition to the SM particle content. The decay processes in question are based on two specific triple-leptoquark interaction topologies that we show in Fig.~\ref{fig:topologies}, where $q$'s and $\ell$'s denote generic quarks and leptons of the SM, while $\langle H \rangle$ stands for a vacuum expectation value of the SM Higgs boson doublet $H$. The scalar leptoquark states $\Delta^Q$, $\Delta^{Q^\prime}$, and $\Delta^{Q^{\prime\prime}}$ in Fig.~\ref{fig:topologies} carry electric charges $Q$, $Q^\prime$, and $Q^{\prime\prime}$, respectively, and can originate, as we discuss later on in detail, from either two or three different leptoquark multiplets. Note that $\kappa$ is a generic cubic parameter of the $\Delta^Q$-$\Delta^{Q^\prime}$-$\Delta^{Q^{\prime\prime}}$ vertex in Fig.~\ref{fig:topologies}, whereas $\lambda$ stands for a dimensionless quartic coupling of the $\Delta^Q$-$\Delta^{Q^\prime}$-$\Delta^{Q^{\prime\prime}}$-$\langle H \rangle$ vertex.   

Both topologies of Fig.~\ref{fig:topologies} have two different realisations. One is with and the other without the contraction with the Higgs boson doublet, where the diagrams that correspond to the former scenario include a vacuum expectation value leg insertion that is rendered in grey in both panels of Fig.~\ref{fig:topologies}. 

Even though there are already several phenomenological studies~\cite{Kovalenko:2002eh,Klapdor-Kleingrothaus:2002rvk,Arnold:2012sd,Hambye:2017qix,Fonseca:2018ehk,Murgui:2021bdy} of the tree-level proton decay topology that is depicted in the left panel of Fig.~\ref{fig:topologies} there is not a single one, to the best of our knowledge, that looks at the one-loop level baryon number violating topology shown in the right panel. We intend to remedy that and, in the process, demonstrate that the one-loop level topology is much more relevant than the tree-level one regardless of the type of the SM charged fermion that propagates in the loop if and when these two topologies coexist. Our analysis is thus applicable whenever $\ell=\ell^\prime$ in Fig.~\ref{fig:topologies}. We also provide a comprehensive list of the leading-order proton decay channels for all non-trivial cubic and quartic contractions involving three scalar leptoquark multiplets that generate triple-leptoquark interactions of interest, where in the latter case one of the scalar multiplets is the Higgs boson doublet of the SM. 

Scalar leptoquark multiplets relevant for our study and the associated couplings are specified in Table~\ref{tab-3LQ}, where we also explicitly denote transformation properties of these multiplets under the SM gauge group $SU(3) \times SU(2) \times U(1)$. The notation that we use in Table~\ref{tab-3LQ} is self-explanatory and closely follows the notation of a contemporary review of the leptoquark phenomenology~\cite{Dorsner:2016wpm}. We suppress both the $SU(3)$ and $SU(2)$ indices in Table~\ref{tab-3LQ} for compactness and opt to show the flavor indices $i,j(=1,2,3)$ instead. We furthermore use $\vec \tau=(\tau_1,\tau_2,\tau_3)$ to denote Pauli matrices and introduce $\vec S_3=(S_3^1,S_3^2,S_3^3)$ for the $SU(2)$ components of the $S_3$ leptoquark multiplet. Throughout this work we consider the scenarios where scalar leptoquark multiplets couple solely to the quark-lepton pairs. If the leptoquark multiplets could couple directly to the quark-quark pairs we assume such interactions to be either suppressed or altogether absent.
\begin{table}[t]
\renewcommand{\arraystretch}{1.4}
\centering
\begin{tabular}{|c|c|}
\hline
Leptoquark multiplets & Yukawa interactions \\ 
\hline
\hline
$R_2= (\mathbf{3},\mathbf{2},7/6) $ & $ - (y^{L}_{R_2} )_{ij\,} \bar{u}_{R \,i} R_2 i \tau_2 L_j  +  (y^{R}_{R_2} )_{ij} \,\bar Q_i R_2 e_{R\,j} +\mathrm{h.c.}$\\[0.1em]
 $\tilde{R}_2 =(\mathbf{3},\mathbf{2},1/6) $ & $ - (y^{L}_{\tilde{R}_2} )_{ij}\, \bar{d}_{R \,i} \tilde{R}_2 i \tau_2 L_j  +\mathrm{h.c.} $\\ $S_1=(\mathbf{ \bar 3},\mathbf{1},1/3) $ & $ (y^{L}_{S_1} )_{ij} \,\bar{Q}^C_i i \tau_2 S_1 L_j  +  (y^{R}_{S_1} )_{ij}\, \bar u_{R\, i}^C S_1 e_{R\,j} +\mathrm{h.c.}$\\[0.1em]
$S_3 =(\mathbf{ \bar 3},\mathbf{3},1/3) $ & $ (y^{L}_{S_3} )_{ij} \,\bar{Q}^C_i i \tau_2 (\vec \tau \cdot \vec S_3) L_j  +\mathrm{h.c.}$\\[0.1em]
$\tilde{S}_1= (\mathbf{ \bar 3},\mathbf{1},4/3) $ & $  (y^{R}_{\tilde{S}_1} )_{ij}\, \bar d_{R\, i}^C \tilde S_1 e_{R\,j} +\mathrm{h.c.}$\\
\hline 
\end{tabular}
\caption{Scalar leptoquark multiplets and their interactions with the SM quark-lepton pairs.}
\label{tab-3LQ} 
\end{table}

The paper is organised as follows. In Sec.~\ref{sec:classification} we study all possible cubic and quartic scalar interactions if one adds to the SM particle content up to three different scalar leptoquark multiplets, specify these interactions at the $SU(3) \times U(1)_\mathrm{em}$ level, and outline main proton decay channels these interactions generate for the two aforementioned decay topologies. In Sec.~\ref{sec:numerical_analysis} we first demonstrate that the one-loop level topology is much more relevant than the tree-level one when it comes to the proton decay signatures. We subsequently present detailed analysis of the effects of the triple-leptoquark interactions on the matter stability within one representative scenario to quantitatively support our claim, where the scenario in question simultaneously features the tree-level topology that yields three-body proton decay and the one-loop level topology that induces two-body proton decay. There we also explicitly show how to extract limit on the energy scale associated with both of these topologies using the most accurate theoretical input and the latest experimental data on partial proton decay lifetimes. We briefly conclude in Sec.~\ref{sec:conclusions}.

\section{Classification}
\label{sec:classification}

We consider extensions of the SM particle content with up to three different scalar leptoquark multiplets generically denoted with $\Delta$, $\Delta^\prime$, and $\Delta^{\prime\prime}$ and study all possible cubic and quartic contractions of the generic forms $\Delta$-$\Delta^\prime$-$\Delta^{\prime\prime}$ and $\Delta$-$\Delta^\prime$-$\Delta^{\prime\prime}$-$H$, respectively, that yield triple-leptoquark interactions $\Delta^Q$-$\Delta^{Q^\prime}$-$\Delta^{Q^{\prime\prime}}$ and $\Delta^Q$-$\Delta^{Q^\prime}$-$\Delta^{Q^{\prime\prime}}$-$\langle H \rangle$. Our aim is to specify the main tree- and one-loop level proton decay channels with topologies of Fig.~\ref{fig:topologies} that can originate from these types of interactions and the associated Yukawa couplings of Table~\ref{tab-3LQ}. Our convention for the transformation properties of the Higgs boson doublet under the SM gauge group $SU(3) \times SU(2) \times U(1)$ is such that $H= (\mathbf{1},\mathbf{2},1/2)$, where we denote its vacuum expectation value with $\langle H \rangle= (0 \;\; v/\sqrt{2})^T$.

If one only demands invariance of the cubic and quartic contractions under the $SU(2) \times U(1)$ part of the SM gauge group, one obtains the following potentially viable terms: $\tilde{R}_2$-$\tilde{R}_2$-$S_1^\ast$~\cite{Kovalenko:2002eh}, $\tilde{R}_2$-$\tilde{R}_2$-$S_3^\ast$~\cite{Kovalenko:2002eh}, $R_2$-$\tilde{R}_2$-$\tilde{S}_1^\ast$~\cite{Kovalenko:2002eh}, $\tilde{R}_2$-$\tilde{R}_2$-$\tilde{R}_2$-$H^\ast$~\cite{Klapdor-Kleingrothaus:2002rvk}, $S_1$-$S_1$-$R_2^\ast$-$H$~\cite{Arnold:2012sd}, $S_1$-$S_3$-$R_2^\ast$-$H$~\cite{Crivellin:2021ejk}, $S_3$-$S_3$-$R_2^\ast$-$H$~\cite{Hambye:2017qix}, $S_1$-$\tilde{S}_1$-$R_2^\ast$-$H^\ast$~\cite{Crivellin:2021ejk}, $S_3$-$\tilde{S}_1$-$R_2^\ast$-$H^\ast$~\cite{Crivellin:2021ejk}, $S_1$-$S_1$-$\tilde{R}_2^\ast$-$H^\ast$~\cite{Arnold:2012sd}, $S_1$-$S_3$-$\tilde{R}_2^\ast$-$H^\ast$~\cite{Crivellin:2021ejk}, and $S_3$-$S_3$-$\tilde{R}_2^\ast$-$H^\ast$~\cite{Crivellin:2021ejk}. 
A thing to note is that it is always possible to replace $S_1$'s with $S_3$'s and vice versa in aforementioned contractions. If one furthermore demands invariance of these contractions under the $SU(3)$ gauge symmetry of the SM, one can demonstrate that the contractions $\tilde{R}_2$-$\tilde{R}_2$-$\tilde{R}_2$-$H^\ast$, $S_1$-$S_1$-$R_2^\ast$-$H$, $\tilde{R}_2$-$\tilde{R}_2$-$S_3^\ast$, and $S_1$-$S_1$-$\tilde{R}_2^\ast$-$H^\ast$ all yield zero~\cite{Crivellin:2021ejk}. These contractions vanish due to a simple fact that they all come out to be symmetric under the exchange of two identical electric charge eigenstates which is in direct conflict with the antisymmetric nature of these contractions in the $SU(3)$ space. Of course, it is always possible to have a new physics scenario where the scalars that transform in the same manner under the SM gauge group are not identical to each other. If that is the case, one would need to revisit those contractions that otherwise would have trivially vanished such as $\tilde{R}_2$-$\tilde{R}_2$-$\tilde{R}_2$-$H^\ast$. 

We summarize all non-trivial cubic and quartic scalar contractions that yield triple-leptoquark interactions in Table~\ref{tab-Lag} at both the $SU(3) \times SU(2) \times U(1)$ and $SU(3) \times U(1)_\mathrm{em}$ levels and specify, to the best of our knowledge, where and when a given contraction has been featured in the literature for the first time. There are two cubic and six quartic contractions, all in all, that generate triple-leptoquark interactions of our interest. The classification presented in Table~\ref{tab-Lag} nicely dovetails an all-encompassing classification of invariant contractions between two scalar leptoquark multiplets and either one or two Higgs boson doublets~\cite{Hirsch:1996qy}. 

Note that the superscript in the second column of Table~\ref{tab-Lag} denotes electric charge $Q$ of leptoquark $\Delta^Q$ in units of electric charge of positron, while $a$, $b$, and $c$ are the leptoquark $SU(3)$ indices. We write $\Delta^{Q\ast}\equiv\Delta^{-Q}$ in the second column of Table~\ref{tab-Lag} for simplicity, where we also define the electric charge eigenstates of $S_3$ leptoquark via $S_3^{1/3}=S_3^3$, $S_3^{4/3}=(S_3^1-iS_3^2)/\sqrt{2}$, and $S_3^{-2/3}=(S_3^1+iS_3^2)/\sqrt{2}$\,.
 \begin{table}[t]
 \resizebox{\textwidth}{!}{
 \renewcommand{\arraystretch}{1.5}
\centering
\begin{tabular}{|cc|c|c|}
\hline
& $SU(3) \times SU(2) \times U(1)$ level & $SU(3) \times U(1)_\mathrm{em}$ level & Ref. \\[0.25em] 
\hline
\hline
$(a)$& $\kappa \tilde{R}_2^T  i \tau_2  \tilde{R}_2 S_1^\ast$& $ - 2 \kappa \epsilon_{abc} \tilde{R}_{2a} ^{-1/3} \,\tilde R_{2b} ^{2/3} \, S_{1c}^{-1/3} $ & \cite{Kovalenko:2002eh}  \\[0.25em]
  $(b)$ & $ \kappa R_2^T i \tau_2  \tilde{R}_2 \tilde{S}_1^\ast $ & $ \kappa \epsilon_{abc} \left(R_{2 a} ^{5/3} \, \tilde{R}_{2b} ^{-1/3} \, \tilde{S}_{1c}^{-4/3}- R_{2 a} ^{2/3} \, \tilde{R}_{2b} ^{2/3} \,\tilde{S}_{1c}^{-4/3} \right)$ & \cite{Kovalenko:2002eh} \\[0.25em]
    $(c)$ & $\lambda H^{\dag} i \tau_2 (\vec \tau \cdot \vec S_3 )^\ast i \tau_2 R_2 S_1^\ast$ & $ \lambda \frac{v}{\sqrt 2}  \epsilon_{abc}  \left( - S_{3a}^{-1/3} \, R_{2b}^{2/3} \, S_{1c}^{-1/3} + \sqrt{2} S_{3a}^{-4/3} \, R_{2b}^{5/3} \, S_{1c}^{-1/3} \right)$ & \cite{Crivellin:2021ejk} \\[0.25em] 
$(d)$ & $ \lambda H^\dag i \tau_2 (\vec \tau \cdot \vec S_3 )^\ast  (\vec \tau \cdot \vec S_3 )^\ast i \tau_2 R_2 $ & $\lambda v \sqrt{2} \epsilon_{abc}  \left(\sqrt{2} S_{3a}^{-1/3} \, S_{3b}^{-4/3} \, R_{2c}^{5/3} -  S_{3a}^{-4/3} \, S_{3b}^{2/3} \, R_{2c}^{2/3}\right)$ & \cite{Hambye:2017qix} \\[0.25em]
  $(e)$ & $\lambda H^T i \tau_2 R_2 S_1^\ast \tilde{S}_1^\ast$ & $-\lambda \frac{v} {\sqrt 2} \epsilon_{abc} R_{2a}^{5/3} \, S_{1b}^{-1/3} \, \tilde{S}_{1c}^{-4/3}$ & \cite{Crivellin:2021ejk} \\[0.25em]
$(f)$ & $\lambda H^T (\vec \tau \cdot \vec S_3 )^\ast i\tau_2 R_2 \tilde{S}_1^\ast$ &$\lambda \frac{v}{\sqrt 2}  \epsilon_{abc}  \left( \sqrt{2} S_{3a}^{2/3} \, R_{2b}^{2/3} \, \tilde{S}_{1c}^{-4/3} + S_{3a}^{-1/3} \, R_{2b}^{5/3} \, \tilde{S}_{1c}^{-4/3}\right)  $ & \cite{Crivellin:2021ejk} \\[0.25em]
 $(g)$ &  $ \lambda H^T(\vec \tau \cdot \vec S_3 )^\ast i \tau_2 \tilde{R}_2 S_1^\ast$ & $\lambda \frac{v}{\sqrt 2}  \epsilon_{abc} \left( \sqrt{2} S_{3a}^{2/3} \, \tilde{R}_{2b}^{-1/3} \, S_{1c}^{-1/3} +S_{3a}^{-1/3} \, \tilde{R}_{2b}^{2/3}  \, S_{1c}^{-1/3}\right)$ & \cite{Crivellin:2021ejk} \\[0.25em]
 $(h)$ & $\lambda H^\dag (\vec \tau \cdot \vec S_3 )^\ast (\vec \tau \cdot \vec S_3 )^\ast  i \tau_2  \tilde{R}_{2} $ & $ \lambda v \sqrt{2}\epsilon_{abc}  
\left(\sqrt{2} S_{3a}^{2/3} \, S_{3b}^{-1/3} \, \tilde{R}_{2c}^{-1/3} + S_{3a}^{-4/3} \, S_{3b}^{2/3} \, \tilde{R}_{2c}^{2/3}\right)$ & \cite{Crivellin:2021ejk}\\
\hline 
\end{tabular}
}
\caption{Cubic and quartic leptoquark multiplet contractions at the $SU(3) \times SU(2) \times U(1)$ level and the associated triple-leptoquark interactions at the $SU(3) \times U(1)_\mathrm{em}$ level.}
\label{tab-Lag}
\end{table}

We can finally specify main proton decay mediating processes for both topologies of Fig.~\ref{fig:topologies} using the Yukawa couplings presented in Table~\ref{tab-3LQ} together with the cubic and quartic interaction terms given in Table~\ref{tab-Lag}. These results are shown in Table~\ref{tab:table-electron} under the assumption that the final states comprise exclusively the first generations of both quarks and charged leptons. We accordingly neglect all proton decay processes induced by the Cabbibo-Kobayashi-Maskawa (CKM) mixing matrix entries, if any. We also write down in Table~\ref{tab:table-electron} relevant operators behind these processes.
\begin{table}[p]
\renewcommand{\arraystretch}{1.65}
\centering
\begin{tabular}{|cc|c|l|l|}
\hline 
& Contractions & Operators & Processes (tree) & Processes (loop)\\ \hline
\hline
\multirow{2}{*}{$(a)$} & \multirow{2}{*}{$\tilde{R}_2$-$\tilde{R}_2$-$S_1^\ast$} &  $ddd\bar{e}\nu\bar{\nu}$ & $p\to \pi^+ \pi^+ e^- \nu \bar{\nu}$ & -- \\ 
& & $ddue\bar{e} \bar{\nu}$ & $p\to \pi^+ e^+ e^- \nu$ &  $p\to \pi^+ \nu$\\  
\hline

\multirow{2}{*}{$(b)$} & \multirow{2}{*}{$R_2$-$\tilde{R}_2$-$\tilde{S}_1^\ast$} & $ddd e \bar{e} \bar{e}$ & $p\to\pi^+\pi^+ e^- e^+ e^-$ & -- \\ 
& & $ddue\bar{e}\bar{\nu}$ & $p\to\pi^+ e^+ e^- \nu$ & $p\to \pi^+ \nu$\\
\hline

\multirow{4}{*}{$(c)$} & \multirow{4}{*}{$S_1$-$S_3$-$R_2^\ast$-$H$} & $ddue\bar{e}{\nu}$ &  $p\to\pi^+ e^+ e^- \bar{\nu}$ & $p\to \pi^+ \bar{\nu}$\\
& & $duu e \nu\bar{\nu}$  & $p\to e^+ \nu \bar{\nu}$ & $p\to \pi^0 e^+$\\
& & $duu e e \bar{e}$ & $p\to e^+ e^+ e^-$ & $p\to \pi^0 e^+$\\
& & $uuu e e \bar{\nu}$& $p \to \pi^- e^+ e^+ \nu$ & --\\
\hline

\multirow{3}{*}{$(d)$} & \multirow{3}{*}{$S_3$-$S_3$-$R_2^\ast$-$H$} & $ddue\bar{e}{\nu}$ & $p\to \pi^+ e^+ e^- \bar{\nu}$ & $p\to \pi^+ \bar{\nu}$\\ 
& & $duu e \nu\bar{\nu}$ & $p \to e^+ \nu \bar{\nu}$ & --\\ 
& & $duu e e \bar{e}$ & $p \to e^+ e^+ e^-$ & $p\to \pi^0 e^+$\\
\hline

\multirow{2}{*}{$(e)$} & \multirow{2}{*}{$S_1$-$\tilde{S}_1$-$R_2^\ast$-$H^\ast$} & $ddue\bar{e}{\nu}$ &  $p\to\pi^+ e^+ e^- \bar{\nu}$ & $p\to \pi^+ \bar{\nu}$\\ 
& &$duu e e \bar{e}$ & $p\to e^+ e^+ e^-$ & $p\to \pi^0 e^+$\\
\hline

\multirow{3}{*}{$(f)$} & \multirow{3}{*}{$S_3$-$\tilde{S}_1$-$R_2^\ast$-$H^\ast$} & $ddue\bar{e}{\nu}$&  $p\to\pi^+ e^+ e^- \bar{\nu}$ & $p\to \pi^+ \bar{\nu}$\\ 
& & $duu e \nu\bar{\nu}$  & $p\to e^+ \nu \bar{\nu}$ & $p\to \pi^0 e^+$\\
& & $duu e e \bar{e}$ & $p\to e^+ e^+ e^-$ & $p\to \pi^0 e^+$\\
\hline

\multirow{4}{*}{$(g)$} & \multirow{4}{*}{$S_1$-$S_3$-$\tilde{R}_2^\ast$-$H^\ast$} & $ddu\nu\bar{\nu}{\nu}$ & $p\to \pi^+ \nu \bar{\nu} \bar{\nu}$ & $p\to \pi^+ \bar{\nu}$\\
& & $ddu e \bar{e}\nu$ &  $p\to\pi^+ e^+ e^- \bar{\nu}$ & $p\to \pi^+ \bar{\nu}$\\
& & $duu e \nu\bar{\nu}$ & $p\to e^+ \nu \bar{\nu}$ & $p\to \pi^0 e^+$\\
& & $duu e e \bar{e}$ & $p\to e^+ e^+ e^-$ & $p\to \pi^0 e^+$\\
\hline
 
\multirow{3}{*}{$(h)$} & \multirow{3}{*}{$S_3$-$S_3$-$\tilde{R}_2^\ast$-$H^\ast$} & $ddu\nu\bar{\nu}{\nu}$ & $p\to \pi^+ \nu \bar{\nu} \bar{\nu}$ & $p\to \pi^+ \bar{\nu}$\\
& & $ddue\bar{e}\nu$ &  $p\to\pi^+ e^+ e^- \bar{\nu}$ & -- \\
& & $duu e \nu\bar{\nu}$  & $p\to e^+ \nu \bar{\nu}$ & $p\to \pi^0 e^+$\\
\hline

\end{tabular}
\caption{List of all non-trivial $\Delta$-$\Delta^\prime$-$\Delta^{\prime\prime}$ and $\Delta$-$\Delta^\prime$-$\Delta^{\prime\prime}$-$H$ contractions, schematic representation of the associated $d=9$ effective operators, and corresponding proton decay channels at both the tree- and one-loop levels. The effective operators in scenarios $(a)$ and $(b)$ conserve $B+L$, while the ones appearing in the remaining scenarios conserve $B-L$, where $B$ and $L$ are baryon and lepton numbers, respectively.}
\label{tab:table-electron}
\end{table}

\section{Phenomenological analysis}
\label{sec:numerical_analysis}
We extract, in this Section, lower limits on the energy scales that are associated with the proton decay signatures due to the presence of triple-leptoquark interactions by using the latest experimental constraints on partial proton decay lifetimes~\cite{ParticleDataGroup:2020ssz} for both the tree-level and one-loop level topologies of Fig.~\ref{fig:topologies}. We denote these energy scales simply with $\Lambda$, where we identify $\Lambda$ with a common scale for the masses of all those scalar leptoquarks that participate in a given baryon-number-violating process under consideration. 

\subsection{Tree- vs.\ one-loop level proton decays}
\label{sec:dim-analysis}
Let us first estimate the expected proton decay widths for the tree- and one-loop level topologies from Fig.~\ref{fig:topologies} to demonstrate that the latter dominates in all instances. We focus, for illustrative purposes, on the processes $p \to e^+ e^+ e^-$ and $p \to \pi^0 e^+$ that are induced for most scenarios in Table~\ref{tab:table-electron}. 

We find the decay rate of the tree-level process, using naive dimensional analysis, to be\footnote{We focus on the scenarios of the type $\Delta$-$\Delta^\prime$-$\Delta^{\prime\prime}$-$H$, but our results also apply to the $\Delta$-$\Delta^\prime$-$\Delta^{\prime\prime}$ scenarios if one replaces $\kappa$ with a product $\lambda v$.}
\begin{equation}
\label{eq:example-p3e}
\Gamma(p\to e^+ e^+ e^-) \simeq \dfrac{m_p}{(16 \pi)^3} \bigg{(}\dfrac{m_p^5 v}{\Lambda^6 }\bigg{)}^2 |\lambda\,y_{ue}^2\, y_{de}|^2\,,
\end{equation}
where $y_{qe}$ denotes generic leptoquark couplings to electrons and valence quarks $(q=u,d)$, and $\Lambda$ stands for the common mass of leptoquarks that are taken to be mass-degenerate. The dependence on $\Lambda$ in Eq.~\eqref{eq:example-p3e} arises from the leptoquark propagators depicted in the left panel of Fig.~\ref{fig:topologies}. 

We can perform an analogous estimate of the decay width of the corresponding one-loop level process $p\to \pi^0 e^ +$ if we note that the loop contribution in Fig.~\ref{fig:topologies} can be schematically seen as an effective diquark coupling that reads
\begin{equation}
\label{eq:diquark-estimate}
y_{u d} \simeq \dfrac{1}{16 \pi^2}\dfrac{m_f v}{\Lambda^2} \lambda\, y_{ue} y_{de}^\ast \,,
\end{equation}
where $m_f$ can be either the valence quark mass $m_q$, or the mass of the lepton running in the loop, i.e., $m_e$ in this case, depending on the specific scenario, as will be discussed below. (See also Appendix~\ref{app:loops} for more details.) The loop-induced proton decay can then be expressed as follows
\begin{equation}
\label{eq:example-ppie}
\Gamma(p\to \pi^0 e^+) \simeq \dfrac{m_p}{16 \pi}  \bigg{(}\dfrac{m_p^2}{\Lambda^2 }\bigg{)}^2 |y_{ud} \, y_{ue}|^2\,.
\end{equation}
By combining Eqs.~\eqref{eq:example-p3e} and~\eqref{eq:example-ppie}, and taking the electron mass as a benchmark value for $m_f$, we find that
\begin{align}
\dfrac{\Gamma(p\to e^+ e^ + e^-)}{\Gamma(p\to \pi^0 e^+)} \simeq  \dfrac{1}{\pi^2}\bigg{(} \dfrac{m_p^3}{m_f \, \Lambda^2}\bigg{)}^2 \simeq  10^{-7} \bigg{(}\dfrac{m_e}{m_f}\bigg{)}^2 \bigg{(}\dfrac{1\,\mathrm{TeV}}{\Lambda}\bigg{)}^{4}\,,
\end{align}
where the dependence on the leptoquark couplings cancels out to the first approximation. 

It is now transparent that proton decays faster through the one-loop level induced two-body process than through the tree-level three-body one. If we also take into account that the experimental limit for the partial lifetime of $p\to \pi^0 e^+$ is roughly one order of magnitude more stringent than the one for the three-body decay such as $p\to e^+ e^+ e^-$~\cite{ParticleDataGroup:2020ssz}, we can conclude with certainty that the loop-induced processes are more sensitive probes of the triple-leptoquark interactions than the tree-level ones. Our estimate is based on the scenario where $m_f= m_e$ and it would be even further exacerbated if the chirality is flipped in the quark lines, leading to $m_f=m_q$, or if heavier leptons are running in the loop. 

We finally opt to show how to accurately perform the extraction of a lower limit on the leptoquark masses that we denote with $\Lambda$ within the framework of scenario $(d)$ that is defined in Tables~\ref{tab-Lag} and \ref{tab:table-electron} for both the tree-level $p \to e^+ e^+ e^-$ decay, and the one-loop level decays $p\to \pi^0 e^+$ and  $p\to \pi^+ \bar{\nu}$. To deduce $\Lambda$ we will eventually set all of the dimensionless couplings to one and focus exclusively on the leptoquark couplings to the first generation of fermions in Secs.~\ref{sec:e-tree} and \ref{sec:e-loop}. 

\subsection{Tree-level leptoquark mediation of $p \to e^- e^+ e^+$}
\label{sec:e-tree}

\begin{figure}[t!]
\centering
\includegraphics[width=0.45\textwidth]{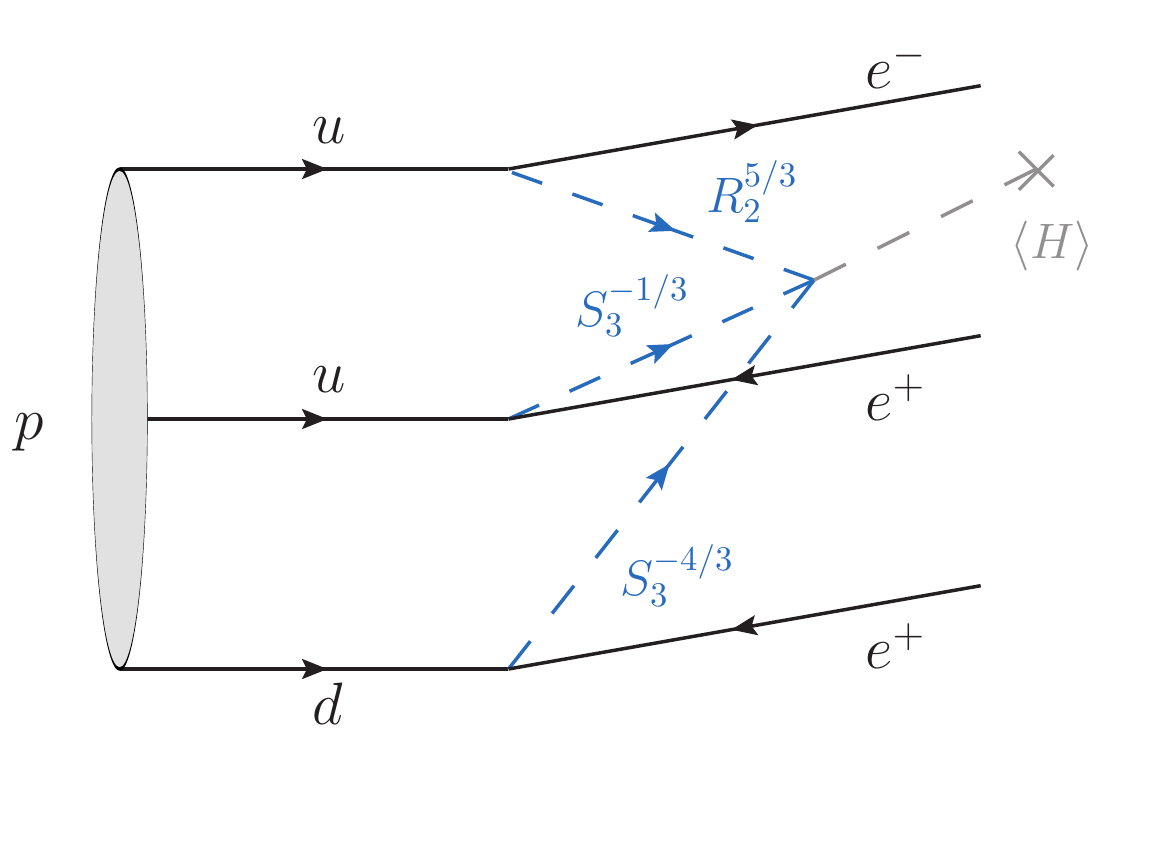}
\caption{Tree-level diagram contributing to the process $p\to e^+ e^+ e^-$ in scenario $(d)$ defined in Table~\ref{tab-Lag}.}
\label{fig:3lq-tree}
\end{figure}

Let us first consider decay amplitude for $p \to e^+ e^+ e^-$ and determine the corresponding decay rate for scenario $(d)$ from Table~\ref{tab-Lag}, i.e., the $S_3$-$S_3$-$R_2^\ast$-$H$ contraction. We, again, focus on the case where leptoquarks couple only to fermions of the first generation. The starting point of our analysis is the $d=9$ effective Lagrangian obtained after integrating out $R_2$ and $S_3$ scalar leptoquarks, see Fig.~\eqref{fig:3lq-tree}, that reads
\begin{align}
 \mathcal L^{(d=9)}_\mathrm{eff} \supset &\sum_{X=L,R} \epsilon_{abc} \, C_{X} \,   ( \bar u^C_a P_L e ) ( \bar d^C_b  P_L e) ( \bar e P_X u_c )+\mathrm{h.c.}\,,
\label{ell-d}
\end{align}
where $a,b,c$ are color indices and the effective coefficients read
\begin{align}
\label{eq:aaa}
C_{L} &= \dfrac{2\sqrt{2} \lambda v}{m_{S_3}^4 m_{R_2}^2} (V^\ast y_{S_3  }^{L} ) y_{S_3}^L (V y_{R_2}^{R} )^\ast \,,\\[0.3em]
\label{eq:bbb}
C_{R} &= -\dfrac{2\sqrt{2} \lambda v}{m_{S_3}^4 m_{R_2}^2}  (V^\ast y_{S_3  }^{L} ) y_{S_3}^L (y_{R_2}^{L} )^\ast\,.
\end{align} 
Note that all flavor indices $i$ and $j$ of Table~\ref{tab-3LQ} are set to $1$ and thus not written out. Also, $V$ in Eqs.~\eqref{eq:aaa} and~\eqref{eq:bbb} is the CKM matrix that, in our convention, resides in the up-type quark sector. We keep it in our calculation for bookkeeping purposes even though it can be treated as an identity matrix in what follows. 

To estimate the decay width for the process $p\to e^+ e^+ e^-$, we use the Fierz relations in Eq.~\eqref{ell-d} which produce the following scalar matrix elements\footnote{The Fierz relations also generate tensor operators such as $(\bar u^C_a \sigma_{\mu\nu} P_L d_b ) ( \bar e  P_L u_c) ( \bar e^C \sigma^{\mu\nu}P_L e )$, but these operators vanish identically due to the properties of the charge-conjugation matrix.}
 \begin{align}
 \begin{split}
 \epsilon_{abc} \langle 0 \vert (\bar u^C_a P_R d_b) P_L u_c \vert p\rangle = \alpha_p P_R u_p\,,\\[0.4em]
 \epsilon_{abc} \langle 0 \vert (\bar u^C_a P_L d_b) P_L u_c  \vert p\rangle = \beta_p P_L u_p\,,
\end{split}
\label{ab-lattice}
\end{align}
where  $u_p$ is the proton spinor, whereas $\alpha$ and $\beta$ are hadronic parameters that have been obtained by numerical simulations of QCD on the lattice~\cite{Aoki:2017puj}. These are
 \begin{align}
 \begin{split}
 \alpha_p&= -0.0144(3)(21)\,\mathrm{GeV}^3\,,\\[0.35em]
 \beta_p&= +0.0144(3)(21)\,\mathrm{GeV}^3\,. 
 \end{split}
 \end{align}

\noindent  We can use this input to write in full generality that
\begin{equation}
\label{eq:dgamma}
\Gamma(p\to e^+e^+e^-) = \dfrac{m_p^5}{6(16\pi)^3}\left(\beta_p^2 |C_L|^2+\alpha_p^2 |C_R|^2\right)\,.
\end{equation}
If we now take $y_{S_3  }^{L}= y_{R_2}^{L}= y_{R_2}^{R}=\lambda = 1$ with $m_{S_3} =  m_{R_2} = \Lambda$ and require that the calculated rate via Eq.~\eqref{eq:dgamma} does not saturate the experimental limit, i.e., $\tau (p \to e^+ e^+ e^-)> 7.93 \times 10^{32}$\,years~\cite{ParticleDataGroup:2020ssz}, we find the following constraint 
\begin{equation}
\label{eq:bound-p3e}
p\to e^+ e^+ e^-:\qquad\Lambda \geq 1.2 \times 10^{2}\,\mathrm{TeV}\,.
\end{equation}
Again, the lower bound on $\Lambda$ in Eq.~\eqref{eq:bound-p3e} corresponds to the energy scale at which the experimental limit for $p \to e^+ e^+ e^-$ is saturated for order one dimensionless couplings and mass-degenerate leptoquarks in scenario $(d)$ from Table~\ref{tab-Lag}. 

\subsection{Loop-level leptoquark mediation of $p\to \pi^0 e^+$ and $p\to \pi^+ \bar{\nu}$}
\label{sec:e-loop}

We now turn our attention to the two-body proton decays $p\to \pi^0 e^+$ and  $p\to \pi^+ \bar{\nu}$ that are induced at the one-loop level via the diagrams in the right panel of Fig.~\ref{fig:topologies}. We, again, consider scenario $(d)$ of Table~\ref{tab-Lag} with the assumption that the leptoquarks couple only to the first generation SM fermions.
 
We start by discussing the main features of the one-loop level decay topology depicted in Fig.~\ref{fig:topologies}. This contribution can be understood as a loop-induced diquark coupling of the leptoquark state $\Delta^{Q^{\prime\prime}}$, which then contributes to the two-body proton decay modes in the usual way, i.e., via $d=6$ operators. However, the $SU(3)$ structure of the one-loop level topology in Fig.~\ref{fig:topologies} imposes important restrictions on the possible external quark states. Since the triple-leptoquark vertex is fully antisymmetric in the $SU(3)$ indices, the one-loop level contributions vanish if the quarks $q$ and $q^\prime$ in Fig.~\ref{fig:topologies} are identical. In other words, the one-loop level contributions are only present if $q$ and $q^\prime$ carry different flavors. 

In the following, we discuss the phenomenology of scenario $(d)$, deferring the details of the one-loop computation for Appendix~\ref{app:loops}.

\begin{figure}[t!]
\centering
\includegraphics[width=0.48\textwidth]{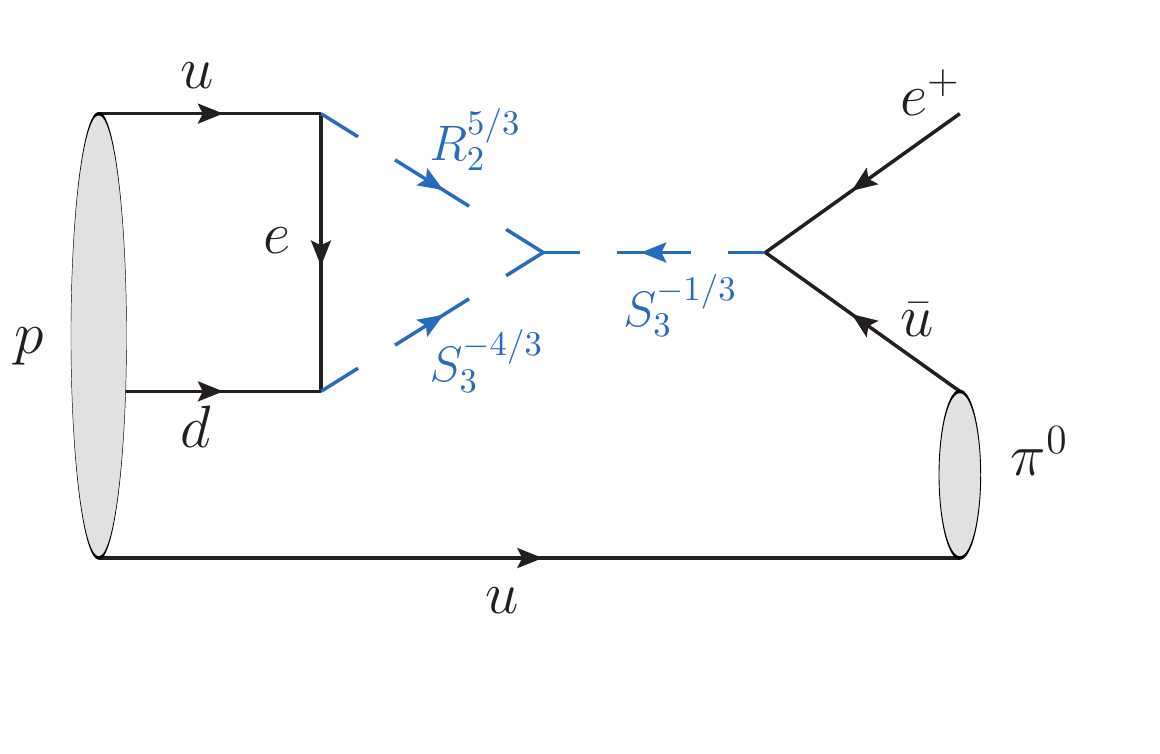}~\qquad~\includegraphics[width=0.48\textwidth]{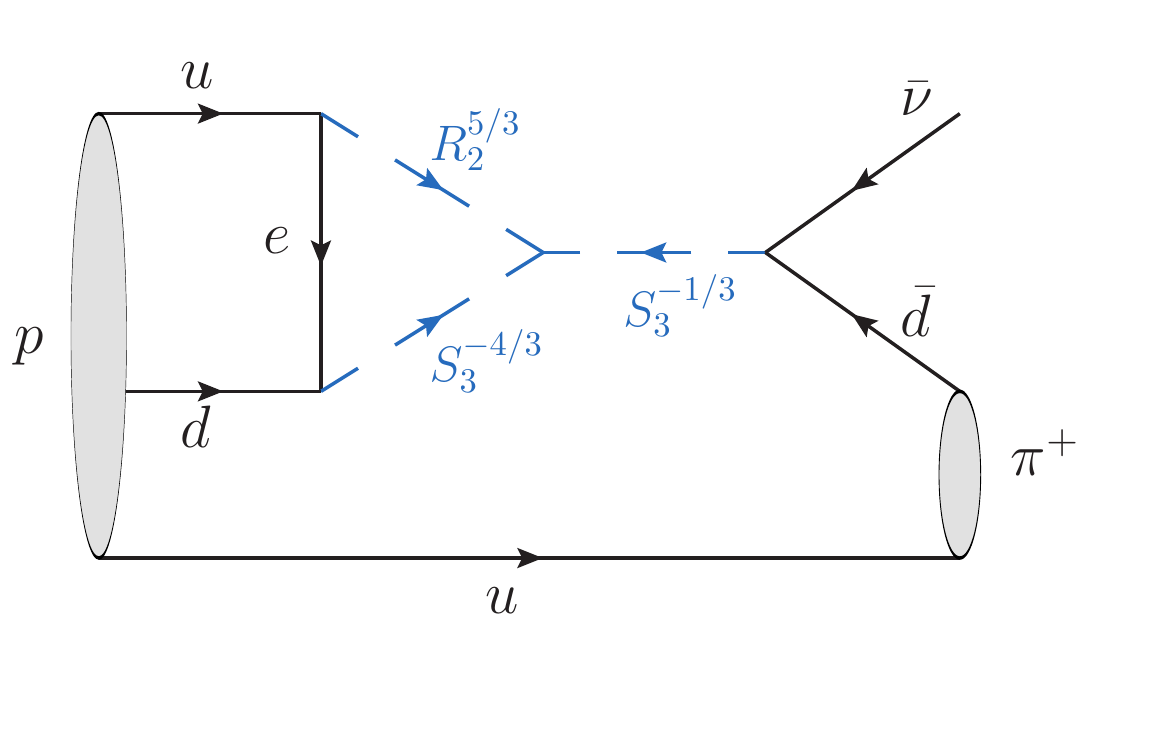}
\caption{One-loop level diagrams contributing to $p\to \pi^0 e^+$ (left panel) and $p\to \pi^+ \bar{\nu}$ (right panel) in scenario $(d)$ of Table~\ref{tab-Lag}.}
\label{fig:3lq-loop}
\end{figure}

\paragraph{\underline{$p\to \pi^0 e^+$}\,:} The most interesting probe of triple-leptoquark interactions is the decay $p\to \pi^0 e^+$ due to the stringent experimental limit $\tau(p\to\pi^0 e^+)^\mathrm{exp}>2.4 \times 10^{34}$\,years~\cite{Super-Kamiokande:2020wjk}. In the scenario we consider, there is only one diagram that contributes to this process at one loop, as depicted in Fig.~\ref{fig:3lq-loop}. We assume that the leptoquark states are degenerate in mass, with $m_{R_2}=m_{S_3} \equiv \Lambda$. After integrating out the leptoquarks, we obtain the effective Lagrangian,
\begin{equation}
\mathcal{L}_\mathrm{eff}^{(d=6)} \supset C^{udeu}_{LL}\,\left(\overline{u}^{\textrm{C}}P_{L}d\right)\left( \overline{e}^{\textrm{C}}P_{L}u\right) + C^{udeu}_{LR}\,\left(\overline{u}^{\textrm{C}}P_{L}d\right)\left( \overline{e}^{\textrm{C}}P_{R}u\right) + \mathrm{h.c.}\,,
\end{equation}

\noindent where the Wilson coefficients, at the scale $\Lambda$, read
\begin{align}
C_{LL}^{udeu} &= \dfrac{\sqrt{2}\lambda }{8\pi^2 } \dfrac{v m_e}{\Lambda^4} (V^\ast y^L_{S_3}) \left[ y^L_{S_3} (V y_{R_2}^R)^\ast +\dfrac{m_d}{4 m_e}y^L_{S_3} (y_{R_2}^L)^\ast \right]\,, \\[0.4em]
C_{LR}^{udeu} &= \dfrac{\lambda }{32\pi^2 } \dfrac{v m_u}{\Lambda^4}(V^\ast y^L_{S_3})^2 (y_{R_2}^L)^\ast \,.
\end{align}

\noindent We, again, omit flavor indices since they are all set to one. The hadronic matrix elements needed to compute $p\to \pi^0 e^+$ can be parameterized in full generality as~\cite{Aoki:2017puj,Aoki:2006ib} 
\begin{equation}
\label{eq:matrix-element}
\left\langle \pi^0 \left|\mathcal{O}^{\Gamma\Gamma'}\right|p\right\rangle=\left[W_0^{\Gamma\Gamma'}(q^2)-\frac{i\slashed{q}}{m_p}W_1^{\Gamma\Gamma'}(q^2)\right]P_{\Gamma'}u_p\,,
\end{equation}

\noindent where $\mathcal{O}^{\Gamma\Gamma'}=\left(\overline{u}^{\textrm{C}}P_{\Gamma}d\right)P_{\Gamma'}u$, with $\Gamma, \Gamma'=R,L$. The proton spinor is, again, denoted by $u_p$, $q$ stands for the momentum exchanged in this transition, and $W_0^{\Gamma\Gamma'}(q^2)$ and $W_1^{\Gamma\Gamma'}(q^2)$ are two independent hadronic form-factors. For the $p\to \pi^0 e^+$ transition, the latter form-factors can be neglected since their contributions are suppressed by $m_e/m_p$. The proton decay width can then be expressed in terms of the $W_0$ form factors as follows,
\begin{equation}
\Gamma(p\to \pi^0 e^+)= \dfrac{m_p}{32\pi} \left(1-\frac{m_\pi^2}{m_p^2}\right)^2\Big{[} (W_0^{LL})^2 |C_{LL}^{udeu}|^2+ (W_0^{RL})^2 |C_{LR}^{udeu}|^2\Big{]} \,,
\end{equation}

\noindent where the electron mass has been neglected. $W_0^{\Gamma\Gamma'}\equiv W_0^{\Gamma\Gamma'}(0)$ have been computed on the lattice and they are predicted, for this specific transition, to be $W_0^{LL}=0.134(5)\,\mathrm{GeV}^2$ and $W_0^{LR}=-0.131(4)\,\mathrm{GeV}^2$~\cite{Aoki:2017puj}.

Using the expressions derived above, and assuming that $y_{S_3}^{L} = y_{R_2}^{R} = y_{R_2}^{L}=\lambda=1$, we obtain that the scale $\Lambda$ for $p\to \pi^0 e^+$ should satisfy
\begin{equation}
\label{eq:bound-ppie}
p\to \pi^0 e^+: \quad \Lambda \geq 1.8 \times 10^{4}\,\mathrm{TeV}\,.
\end{equation}
This limit is about two orders of magnitude more stringent than the limit derived from the tree-level proton decay $p\to e^+e^+e^-$ presented in Eq.~\eqref{eq:bound-p3e}. This is in agreement with our initial estimate from Sec.~\ref{sec:dim-analysis}.

\paragraph{\underline{$p\to \pi^+ \bar{\nu}$}\,:} Another interesting proton decay mode is $p\to \pi^+ \bar{\nu}$, where the experimental limit on this partial decay lifetime is currently at $\tau(p\to\pi^+ \bar{\nu})^\mathrm{exp}>3.9 \times 10^{32}$\,years~\cite{ParticleDataGroup:2020ssz}. This process can be induced by the right diagram in Fig.~\ref{fig:3lq-loop}, which contributes to the effective Lagrangian
\begin{equation}
\mathcal{L}_\mathrm{eff}^{(d=6)} \supset C^{ud\nu d}_{LL}\,\left(\overline{u}^{\textrm{C}}P_{L}d\right)\left( \overline{\nu}^{\textrm{C}}P_{L}d\right) + C^{ud\nu d}_{RL}\,\left(\overline{u}^{\textrm{C}}P_{R}d\right)\left( \overline{\nu}^{\textrm{C}}P_{L}d\right) + \mathrm{h.c.}\,,
\end{equation}

\noindent with the Wilson coefficients 
\begin{align}
C_{LL}^{ud\nu d} &= -\dfrac{\sqrt{2}\lambda }{8\pi^2 } \dfrac{v m_e}{\Lambda^4} (y^L_{S_3})^2 \left[(V y_{R_2}^R)^\ast +\dfrac{m_d}{4 m_e}(y_{R_2}^L)^\ast \right]\,, \\[0.4em]
C_{RL}^{ud\nu d} &=-\dfrac{\lambda }{32\pi^2 } \dfrac{v m_u}{\Lambda^4}(V^\ast y^L_{S_3})^2 (y_{R_2}^L)^\ast \,.
\end{align}

\noindent We, once again, assume degenerate leptoquark masses $m_{R_2} = m_{S_3} \equiv \Lambda$ and omit all flavor indices. By combining the isospin relation
\begin{equation}
\langle \pi^+ \vert \left(\overline{u}^{\textrm{C}}P_{\Gamma}d\right)P_{\Gamma'}d\vert p \rangle=\sqrt{2}\langle \pi^0 \vert \left(\overline{u}^{\textrm{C}}P_{\Gamma}d\right)P_{\Gamma'}u \vert p \rangle\,,
\end{equation}

\noindent with Eq.~\eqref{eq:matrix-element}, we find that
\begin{equation}
\Gamma(p\to \pi^+ \nu)= \dfrac{m_p}{16\pi} \left(1-\frac{m_\pi^2}{m_p^2}\right)^2\Big{[} (W_0^{LL})^2 |C_{LL}^{ud\nu d}|^2+ (W_0^{RL})^2 |C_{RL}^{ud\nu d}|^2\Big{]} \,.
\end{equation}

\noindent If we again assume that $y_{S_3}^{L} = y_{R_2}^{L} = y_{R_2}^{R} = \lambda=1$, we find that the scale $\Lambda$ should satisfy the following limit to be consistent with the current experimental input
\begin{equation}
\label{eq:bound-ppinu}
p\to \pi^+ \bar{\nu}\,: \qquad \Lambda \geq 1.2 \times 10^{4}\,\mathrm{TeV}\,.
\end{equation}
This constraint on $\Lambda$ is slightly weaker than the one quoted in Eq.~\eqref{eq:bound-ppie} for $p\to \pi^0 e^+$ due to a less stringent experimental limit but it is still much more relevant than the one derived from three-body proton decay $p\to e^+e^+e^-$ in Eq.~\eqref{eq:bound-p3e}.

\section{Conclusions}
\label{sec:conclusions}

We study a phenomenological impact of triple-leptoquark interactions on proton stability for two different decay topologies under the assumption that scalar leptoquarks of interest couple solely to the quark-lepton pairs. The first topology has the tree-level structure and it yields three-body proton decays at the leading order. The other topology is of the one-loop nature and it generates two-body proton decay processes instead. The tree-level topology has been analysed in the literature before in the context of baryon number violation while the one-loop level one has not been featured in any scientific study to date.

We demonstrate that it is the one-loop level topology that is producing more stringent bounds on the scalar leptoquark masses of the two, if and when they coexist, thus rendering the extraction of limits using the tree-level topology processes redundant. To quantitatively support our claim we extract a lower limit on the mass scale $\Lambda$ that is associated with the leading order proton decay signatures for both topologies within one particular scenario using the latest theoretical and experimental input. We show that the limit on this scale for the one-loop level process $p\to \pi^0 e^+$ reads $\Lambda \geq 1.8\times 10^{4}$\,TeV when the charged lepton in the loop is an electron. The corresponding limit for the tree-level topology process $p \to e^+ e^+ e^-$ is $\Lambda \geq 1.2 
\times 10^{2}$\,TeV. To generate these limits we identify scale $\Lambda$ with a common scale for the masses of all those leptoquarks that participate in a given baryon number violating process and set all of the dimensionless couplings to one under the assumption that leptoquarks couple solely to the first generation SM fermions.

We also specify the most prominent proton decay signatures due to the presence of all non-trivial cubic and quartic contractions involving three scalar leptoquark multiplets, where in the latter case one of the scalar multiplets is the SM Higgs doublet. 


\section*{Acknowledgements}

The authors would like to thank Damir Be\v{c}irevi\'c, Admir Greljo, Nejc~Ko\v{s}nik, and Luc Schnell for useful discussions and valuable input. S.\ F.\ and I.\ D.\ acknowledge the financial support from the Slovenian Research Agency (research core funding No. P1-0035). I.\ D.\ also acknowledges support of COST Action CA16201. This project has received support from the European Union’s Horizon 2020 research and innovation programme under the Marie Skłodowska-Curie grant agreement N$^\circ$~860881-HIDDeN.

\appendix 

\section{Explicit loop computation}
\label{app:loops}

We provide, in this Appendix, general expressions for the loop diagrams depicted in the right panel of Fig.~\ref{fig:topologies}. For simplicity, we work in the broken electroweak phase, and we consider three leptoquark mass eigenstates $\Delta^Q$, $\Delta^{Q^\prime}$ and $\Delta^{Q^{\prime\prime}}$, where the superscripts denote the electric charges of each state, which satisfy $Q+Q^\prime+Q^{\prime\prime}=0$. These states can originate from either two or three leptoquark multiplets, as it is demonstrated in Table~\ref{tab-Lag}.

We write the triple-leptoquark interaction as follows.
\begin{align}
\label{eq:app-mix}
\mathcal{L}_\mathrm{scalar} \supset \lambda\, v\, \varepsilon_{abc} \, \Delta_a^Q\,\Delta_b^{Q^\prime}\,\Delta_c^{Q^{\prime\prime}} +\mathrm{h.c.}\,,
\end{align}
where $a,b,c$ are color indices, and the coupling $\lambda$ can be easily identified for each of the scenarios in Table~\ref{tab-Lag}. We assume that the leptoquarks $\Delta^Q$ and $\Delta^{Q^\prime}$ carry fermion numbers $F=0$ and $F=2$, respectively, and that they have the following Yukawa interactions,
\begin{align}
\label{eq:app-yuk}
\mathcal{L}_\mathrm{yuk.} \supset  \overline{q} \left( y_{R} P_R + y_{L} P_L \right) \ell  \,\Delta^{Q} + \overline{q^{\prime C}}
 \left( y_{R}^\prime P_R + y_{L}^\prime P_L \right) \ell  \,\Delta^{Q^{\prime}\ast} +\mathrm{h.c.}\,,
\end{align}

\noindent in addition to the Yukawa couplings of $\Delta^{Q^{\prime\prime}}$ which are not explicitly written out. In Eq.~\eqref{eq:app-yuk}, $\ell$ is a generic lepton, and $q$ and $q^\prime$ stand for two distinct quarks, with electric charges satisfying $Q=Q_q - Q_\ell$ and $Q^\prime=-Q_{q^\prime}-Q_\ell$. The Yukawa couplings $y_{L}$ and $y_{R}$ for each scenario can be matched to Table~\ref{tab-3LQ} after expanding the leptoquark multiplets. Note, also, that color and flavor indices are not explicitly written in Eq.~\eqref{eq:app-yuk}.

The loop diagram in Fig.~\ref{fig:topologies} corresponds to a loop-induced diquark coupling of the $\Delta^{Q^{\prime\prime}}$ leptoquark,
\begin{align}
\label{eq:app-diquark}
\mathcal{L}_\mathrm{qq^\prime} = \varepsilon_{abc}\,\overline{q_a^C} \left(y_{q q^\prime}^L P_L + y_{q q^\prime}^L P_R \right)q_b^\prime \,\Delta_c^{Q^{\prime\prime}} +\mathrm{h.c.}\,,
\end{align}

\noindent where $y_{q q^\prime}^L$ and $y_{q q^\prime}^R$ can be fully expressed in terms of the couplings defined above. To perform this computation, it is useful to expand the loop amplitude in the external momenta before integration~\cite{Smirnov:1994tg}. We find that
\begin{align}
y_{q q^\prime}^L &= \dfrac{\lambda v}{16 \pi^2 m_\Delta^2} \left( m_\ell \, y_L^\prime y_R^\ast - \dfrac{m_q}{4} y_R^\prime y_R^\ast - \dfrac{m_{q^\prime}}{4} y_L^\prime y_L^\ast \right) \,, \\[0.4em]
y_{q q^\prime}^R &= \dfrac{\lambda v}{16 \pi^2 m_\Delta^2} \left( m_\ell \, y_R^\prime y_L^\ast - \dfrac{m_q}{4} y_L^\prime y_L^\ast - \dfrac{m_{q^\prime}}{4} y_R^\prime y_R^\ast  \right) \,,
\end{align}
where degenerate leptoquark masses are assumed, i.e., $m_{\Delta^Q} =m_{\Delta^{Q^\prime}} =m_{\Delta^{Q^{\prime\prime}}} \equiv m_\Delta $\,. The terms proportional to $m_\ell$  and $m_{q^{(\prime)}}$  correspond to a chirality-flip in the internal lepton and external quark lines, respectively. From these expressions, it is straightforward to derive the triple-leptoquark interaction contributions for any of the scenarios listed in Table~\ref{tab-Lag} and for any of the loop processes collected in Table~\ref{tab:table-electron}.

\section{$d=9$ effective operators}
\label{app:d9ope}

We collect, in this Appendix, the $d=9$ effective Lagrangians that have been used to generate the entries in Table~\ref{tab:table-electron}. These are
\begin{subequations}
\label{abcdefgh}
\begin{align}
\label{ea}
\mathcal L_{(a)} & \supset  \frac{2 \kappa \epsilon_{abc}}{ m_{\tilde R_2}^4 m_{S_1}^2}  (y_{\tilde R_2}^{L} )_{1j}^\ast ( \bar \nu_L^j  d_{R\,a}) \,(y_{\tilde R_2}^{L} )_{1k}^\ast( \bar e_L^k    d_{R\,b} )\\
&\times \Big{[} (V^\ast y_{S_1}^L )_{1i} (\bar u^C_{L\,c}   e_L^i )-(y_{S_1}^L )_{1i} ( \bar d^C_{L\,c} \nu_L^i) 
+ (y_{S_1}^R )_{1i}(\bar u^C_{R\,c}  e_R^i  )\Big{]} +\mathrm{h.c.}\,, \nonumber
\\[0.65em]
\label{eb}
\mathcal L_{(b)} & \supset   \frac{\kappa \epsilon_{abc}}{ m_{ \tilde S_1}^2 m_{R_2}^2  m_{\tilde R_2}^2} 
\bigg{\lbrace} \Big{[} (V y_{R_2}^R)^\ast_{1j} ( \bar e^j_R  u_{L\,a} ) -(y_{R_2  }^{L})^ \ast_{1j} ( \bar e^j_L  u_{R\,a} )\Big{]} \, (y_{\tilde R_2  }^{L } )_{1k}^\ast ( \bar \nu^k_L d_{R\,b})\\
& +\Big{[}(y_{R_2  }^{L})^ \ast_{1j} ( \bar  \nu^j_L u_{R\,a}) + (y_ {R_2  }^R)^\ast_{1j} (\bar e_R^j d_{L\,a}) \Big{]}(y_{\tilde R_2  }^L)_{1k}^\ast ( \bar e^k_L   d_{R\,b})\bigg{\rbrace}\, (y_{\tilde S_1} ^R)_{1i}( \bar d^C_{L\,c} e^i_R ) +\mathrm{h.c.}\nonumber \,,
\\[0.65em]
 \label{ec}
\mathcal L_{(c)} & \supset   \frac{\lambda \epsilon_{abc} v}{\sqrt{2} m_{ S_3}^2 m_{R_2}^2  m_{S_1}^2}  \Big{[}(V^\ast y_{S_1}^L )_{1i} (\bar u_{L\,c}^C  e_L^i) - (y_{S_1}^L)_{1i}  (\bar d^C_{L\,c} \nu_L^i ) +  (y_{S_1}^R)_{1i} (\bar u_{R\,c}^C e_R^i) \Big{]} \\
&\times \bigg{\lbrace}  \Big{[} (V^\ast y_{S_3}^L )_{1j} (\bar u^C_{L\,a} e^j_L ) +(y_{S_3}^L )_{1j} (\bar d^C_{L\,a} \nu^j_L )\Big{]}  \Big{[} (y_{R_2 }^L )^\ast_{1k} (\bar \nu^k_L u_{R\,b})+(y_{R_2 }^R )_{1k}^\ast (\bar e^k_R  d_{L\,b}) \Big{]} \nonumber
\\
&- \sqrt 2  (y_{S_3}^L  )_{1j}  (\bar d^C_{L\,a}  e_L^j) \Big{[} (V y_{R_2}^R)^\ast_{1k} ( \bar e^k_R  u_{L\,b} ) -(y_{R_2  }^{L})^ \ast_{1k} ( \bar e^k_L  u_{R\,b} )\Big{]}  \bigg{\rbrace}+\mathrm{h.c.}\,,  \nonumber
\\[0.65em]
\label{ed}
\mathcal L_{(d)} & \supset  \frac{2\sqrt 2\lambda \epsilon_{abc} v}{ m_{S_3}^4 m_{R_2}^2} 
( y_{S_3}^L )_{1j}( \bar   d^C_{L\,a} e_L^j)\\
&\times\bigg{\lbrace}  \Bigl.  (V^\ast y_{S_3}^L )_{1k} (\bar u_{L\,b}^C \nu^k_L)
\Big{[} (y_{R_2}^L )_{1i}^\ast  (\bar \nu^i_L u_{R\,c})  + ( y_{R_2}^R)_{1i}^\ast( \bar e^i_R d_{L\,c} )\Big{]} \nonumber \\
&+\Big{[} (y_{S_3  }^{L} )_{1k}( \bar   d^C_{L\,b} \nu^k_L ) +(V^\ast y_{S_3  }^{L} )_{1k}  ( \bar u^C_{L\,b} e_L^k) \Big{]}\Bigr.\Big{[}  (y_{R_2}^L)_{1i}^\ast ( \bar e^i_L u_{R\,c} )  -(Vy_{R_2}^R )_{1i}^\ast ( \bar e^i_R u_{L\,c} ) \Big{]}\bigg{\rbrace}+\mathrm{h.c.}\,,\nonumber
\\[0.65em]
\label{ee}
\mathcal L_{(e)} &\supset  \frac{-\lambda \epsilon_{abc} v}{\sqrt{2} m_{ \tilde S_1}^2 m_{R_2}^2  m_{S_1}^2} (y_{\tilde S_1  }^{R  } )_{1i} ( \bar d^C_{R\,c} e_R^i ) 
\Big{[}  (y_{R_2  }^L)^{ \ast}_{1j}  ( \bar e^j_L    u_{R\,a}  ) -(Vy_{R_2}^R )_{1j}^\ast ( \bar e^j_R u_{L\,a} )\Big{]} \\
& \times \Big{[} (V^\ast y_{S_1}^L )_{1k}  (\bar u^C_{L\,b}  e_L^k) + (y_{S_1}^R)_{1k} (\bar u^C_{R\,b}  e_R^k ) - (y_{S_1}^L)_{1k}  (\bar d^C_{L\,b}  \nu_L^k )\Big{]} +\mathrm{h.c.}\,,\nonumber
\\[0.65em]
\label{ef}
\mathcal L_{(f)} &\supset  \frac{\lambda \epsilon_{abc} v}{\sqrt{2} m_{R_2}^2  m_{S_3}^2  m_{ \tilde S_1}^2 } (y_{\tilde S_1  }^{R  })_{1i} (\bar d^C_c P_R e^i)\\
&\times\bigg{\lbrace} 2( V^\ast y_{S_3}^L )_{1j} (\bar u^C_{L\,a}  \nu_L^j)  \Big{[}(y_{R_2  }^L)^{\ast}_{1k}(\bar  \nu^k_L u_{R\,b})   + (y_{R_2}^R)_{1k}^\ast(\bar e^k_R  d_{L\,b})  \Big{]} \Bigr. \nonumber
\\
&+ \Bigl.\Big{[} (y_{S_3}^L  )_{1j} (\bar d^C_{L\,a}  \nu_L^j ) + (V^\ast  y_{S_3}^L)_{1j} (\bar u_{L\,a}^C e^j_L) \Big{]} \Big{[}(y_{R_2}^L)^\ast_{1k} (\bar e^k_L u_{R\,b})   -(Vy_{R_2}^R )_{1k}^\ast ( \bar e^k_R u_{L\,b} )\Big{]}\bigg{\rbrace}
+\mathrm{h.c.}\,,\nonumber
\\[0.65em]
 \label{eg}
\mathcal L_{(g)} &\supset  \frac{\lambda \epsilon_{abc} v}{\sqrt{2} m_{ S_3}^2 m_{\tilde R_2}^2  m_{S_1}^2} \Big{[} (V^\ast  y_{S_1}^L)_{1i}  (\bar u_{L\,c}^C e_L^i)  -(y_{S_1}^L )_{1i} (\bar d^C_{L\,c} \nu^i_L )  +  (y_{S_1}^R)_{1i}  (\bar u_{R\,c}^C e_R^i) \Big{]}\\
&\times\bigg{\lbrace} \Big{[}  ( y_{S_3 }^L)_{1j} (\bar d^C_{L\,a}\nu_L^j) +  (  V^\ast y_{S_3}^L)_{1j}  (\bar u^C_{L\,a}  e_L^j)\Big{]} (y_{\tilde R_2 }^L)^\ast_{1k} (\bar e^k_L  d_{R\,b} ) \Bigr. \nonumber\\
&+2  ( V^\ast y_{S_3}^{L })_{1j} (\bar u^C_{L\,a}  \nu_L^j )\, (y_{\tilde R_2 }^L)^\ast_{1k}  (\bar \nu^k_L  d_{R\,b})  \bigg{\rbrace} +\mathrm{h.c.}\,, \nonumber
 \\[0.65em]
  \label{eh}
\mathcal L_{(h)} &\supset  \frac{2 \sqrt{2} \lambda \epsilon_{abc} v}{ m_{ S_3}^4 m_{\tilde R_2}^2 } 
\bigg{\lbrace}  \Bigl. 
  ( y_{S_3}^L)_{1j} (\bar d^C_{L\,a} e_L^j)\, ( V^\ast y_{S_3}^L)_{1k} (\bar u_{L\,b}^C  \nu^k_L)\, (y_{\tilde R_2}^L )^\ast_{1i} (\bar e^i_L  d_{R\,c})\\
& -(y_{\tilde R_2 }^L)^\ast_{1i}  (\bar \nu^i_L  d_{R\,c})\,( V^\ast y_{S_3}^{L })_{1j} (\bar u^C_{L\,a} \nu_L^j) \Big{[} (y_{S_3})_{1k}^L (\bar d^C_{L\,b} \nu_L^k ) 
 + (V^\ast y_{S_3}^L )_{1k} (\bar u^C_{L\,b} e^k_L)\Big{]}\nonumber\bigg{\rbrace}+\mathrm{h.c.}\,,
\end{align}
\end{subequations}
where $V$ is the CKM matrix that, in our convention, resides in the up-type quark sector. The SM charged fermions are thus given as the mass eigenstates in Eqs.~\eqref{abcdefgh} whereas the neutrinos are in the flavor eigenstate basis. For simplicity, we only keep the terms that contain the up ($u$) and  down ($d$) quarks since these are the most relevant ones for the proton-decay modes we consider.


\end{document}